\documentclass{llncs}

\usepackage{amssymb}
\usepackage{amsmath}
\begin{document}



\title{Speeding-up Dynamic Programming with Representative Sets \thanks{The third author is supported by the NWO project 'Space and Time Efficient Structural Improvements of Dynamic Programming Algorithms'.}}

\subtitle{An Experimental Evaluation of Algorithms for Steiner Tree on Tree Decompositions}

\titlerunning{Dynamic Programming with Representative Sets }

\author{Stefan Fafianie\inst{1}
\and Hans L. Bodlaender\inst{1}
\and Jesper Nederlof\inst{1}}

\institute{Utrecht University, The Netherlands\\
\email{S.Fafianie@uu.nl; 
H.L.Bodlaender@uu.nl; J.Nederlof@uu.nl}}
\maketitle

\begin{abstract}
Dynamic programming on tree decompositions is a frequently used approach to
solve otherwise intractable problems on instances of small treewidth. 
In recent work by Bodlaender et al. \cite{BodlaenderCKN12}, it was shown that for many connectivity problems,
there exist algorithms that use time, linear in the number of vertices, and 
single exponential in the width of the tree decomposition that is used.
The central idea is that it suffices to compute representative sets, and these
can be computed efficiently with
help of Gaussian elimination.

In this paper, we give an experimental evaluation of this technique for the 
\textsc{Steiner Tree} problem. A comparison of the classic dynamic programming
algorithm and the improved dynamic programming algorithm that employs the
table reduction shows that the new approach gives significant improvements
on the running time of the algorithm and the size of the tables computed by the dynamic programming
algorithm, and thus that the rank based approach from
Bodlaender et al. \cite{BodlaenderCKN12} does not only give significant theoretical improvements but
also is a viable approach in a practical setting, and showcases the potential of
exploiting the idea of representative sets for speeding up dynamic programming algorithms.

\end{abstract}

\section{Introduction}
The notion of treewidth 
provides us with a method of solving many $\mathcal{NP}$-hard problems by means of
dynamic programming algorithms on tree decompositions of graphs, resulting in
algorithmic solutions which are fixed-parameter tractable in the treewidth of the
input graph. For many problems, this gives algorithms that are linear in
the number of vertices $n$, but at least exponential in the width of the tree
decomposition on which the dynamic programming algorithm is executed. 
The dependency of the running time on the width of the tree decomposition
has been a point of several investigations. For many problems, algorithms
were known whose running time is single exponential on the width,
see e.g., \cite{TelleP93}. A recent breakthrough was obtained
by Cygan et al.~\cite{CyganNPPRW11} who showed for several {\em
connectivity} problems, including {\sc Hamiltonian Circuit},
{\sc Steiner Tree}, {\sc Connected Dominating Set} (and many other problems)
that these can be solved in time, single exponential in the width, but at the
cost of introducing randomization and an additional factor in the running time that
is polynomial in $n$. Very recently, Bodlaender et al.~\cite{BodlaenderCKN12}
introduced a new technique (termed the {\em rank based approach})
that allows algorithms for connectivity problems that are
(i) deterministic, (ii) can handle weighted vertices, and (iii) have a running time
of the type $O(c^k n)$ for graphs with a given tree decomposition of width $k$
and $n$ vertices, i.e., the running time is single exponential in the width, and
linear in the number of vertices. 

The main ideas of the rank based approach are the following. (Many
details are abstracted away in the discussion below. See \cite{BodlaenderCKN12}
for more details.)
Suppose
we store during dynamic programming a table $T$ with each entry giving
the characteristic of a partial solution. If we have an entry $s$ in $T$,
such that for each extension of $s$ to a `full solution', $s\cdot t$, there
is an other entry$s'\neq s$ in $T$, that can be extended in the same way to a full solution
$s' \cdot t$, and solution $s'\cdot t$ has a value that is as least as good as the
value of $s\cdot t$, then $s$ is not needed for obtaining an optimal solution,
and we can delete $s$ from $T$. This idea leads to the notion of
{\em representativity}, pioneered by Monien in 1985 \cite{Monien85}.
Consider the matrix $M$ with rows indexed by partial solutions, and
columns indexed by manners to extend partial solution, with a 1.
if the combination gives a full solution, and a 0 otherwise.
A table $T$ corresponds to a set of rows in $M$, with a value
associated to each row. (E.g., for the {\sc Steiner Tree} problem,
a row corresponds to the characteristic of a forest in a subgraph,
and the value is the sum of the edges in the forest.)
It is not hard to see that a maximal subset of linear independent rows of minimal cost
(in case of minimization problems, and of maximal value in case of
maximization problems) forms a representative set. 
Now, if we have an explicit basis of $M$ (the characteristics of the
columns in a maximal set of independent columns in $T$)
and $M$ has `small' rank, then we can find a `small' representative
set efficiently, just by performing Gaussian elimination on a submatrix of $M$.
Now, for many connectivity problems, including {\sc Steiner Tree},
{\sc Feedback Vertex Set}, {\sc Long Path}, {\sc Hamiltonian Circuit},
{\sc Connected Dominating Set}, the rank of this matrix $M$ when
solving these problems on a tree decomposition is single exponential
in the width of the current bag. This leads to the improved dynamic
programming algorithm: interleave the steps of the existing DP algorithm
with computing representative sets by computing the submatrix of $M$
and then carrying out Gaussian elimination on this submatrix.

The notion of representative sets was pioneered by Monien in 1985
\cite{Monien85}. Using the well known two families theorem by Lov\'{a}sz \cite{Lovasz77},
it is possible to obtain efficient FPT algorithms for several other
problems \cite{Marx09,FominLS13}. Cygan et al.~\cite{CyganKN12}
give an improved bound on the rank as function of the width of
the tree decomposition for problems on finding cycles and paths
in graphs of small treewidth, including {\sc TSP}, {\sc Hamiltonian Circuit},
{\sc Long Path}.

In this paper, we perform an {\em experimental evaluation} of the rank based approach,
targeted at the {\sc Steiner Tree} problem, i.e., we discuss an implementation
of the algorithm, described by Bodlaender et al.~\cite{BodlaenderCKN12} for the
{\sc Steiner Tree} problem and its performance. We test the algorithm on
a number of graphs from a benchmark for Steiner Tree, and some randomly generated
graphs. The results of our experiments are very positive: the new algorithm is
considerably faster compared to the classic dynamic programming algorithm,
i.e., the time that is needed to reduce the tables with help of Gaussian elimination
is significantly smaller than the gain in time caused by the fact that tables are
much smaller.

The \textsc{Steiner Tree} problem (of which \textsc{Minimum Spanning Tree} is a special case) is a 
classic $\mathcal{NP}$-hard problem which was one of Karp's original 21 $\mathcal{NP}$-complete problems \cite{Karp72}. 
Extensive overviews on this problem and algorithms for it can be found in \cite{HwangRW92,Winter87}. 
Applications of \textsc{Steiner Tree} include electronic design automation, very large scale integration (VSLI) of circuits and wire routing. 
In this paper we consider the weighted variant, i.e., edges have a weight, and
we want to find a Steiner tree of minimum weight.
It is well known that {\sc Steiner Tree} can be solved in linear time for graphs of bounded treewidth. 
In 1983, Wald and Colbourn \cite{WaldC83} showed this for graphs of treewidth two. For larger fixed values of
$k$, polynomial time algorithms are obtained as consequence of a general characterization by Bodlaender~\cite{Bodlaender87} and linear time algorithms are obtained as consequence of extensions of Courcelles theorem,
by Arnborg et al.~\cite{ArnborgLS91} and Borie et al.~\cite{BoriePT92}. In 1990, Korach and Solel~\cite{KorachS90}
gave an explicit linear time algorithm for {\sc Steiner Tree} on graphs of bounded treewidth.
Inspection shows that the running time of this algorithm is $O(2^{O(k\log k)} n)$; $k$ denotes the width
of the tree decomposition. We call this
algorithm the {\em classic} algorithm. Recently, Chimani et al.~\cite{ChimaniMZ12} gave an improved
algorithm for {\sc Steiner Tree} on tree decompositions that uses $O(B^2_{k+1}\cdot k \cdot n)$ time,
where the Bell number $B_i$ denotes the number of partitions of an $i$ element set.
Our description of the classic algorithm departs somewhat from
the description in Korach and Solel \cite{KorachS90}, but the underlying technique is essentially the same.
We have chosen not to use the coloring schemes from Chimani et al.~\cite{ChimaniMZ12}, but instead 
use hash tables to store information.
Wei-Kleiner \cite{WeiKleiner13} gives a tree decomposition based algorithm for Steiner tree, that 
particularly aims at instances with a small set of Steiner vertices.

In this paper, we compare three different algorithms:
\begin{itemize}
\item The {\em classic} dynamic programming algorithm (CDP), see the discussion above. On a nice tree decomposition,
we build for each node $i$ a table. Tables map partitions of subsets of $X_i$ to values,
characterizing the minimum weight of a `partial solution' that has this partition of a subset
as `fingerprint'. 
\item RBA: To the classic dynamic programming algorithm, we add a step where we
apply the {\em reduce} algorithm from \cite{BodlaenderCKN12}. With help of Gaussian elimination
on a specific matrix (with rows corresponding to entries in the DP table, columns
corresponding to a `basis of the fingerprints of ways of extending partial solutions to
Steiner trees', and values 1, if the extension of the column applied to the entry of the row
gives a Steiner tree and 0 otherwise), we delete some entries from the table. It can be
shown that deleted entries are not needed to obtain an optimal solution, i.e., the
step does not affect optimality of the solution. This elimination step is performed each
time after the DP algorithm has computed a table for a node of the nice tree decomposition.
\item RBC: Similar as RBA, but now the elimination step is only performed for `large'
tables, i.e., tables where the theory tells us that we will delete at least one entry when
we perform the elimination step.
\end{itemize}

The experiments show that both RBA and RBC are both preferable over
the classic algorithm, with RBC outperforming RBA in all cases. Thus, applying
the reduce algorithm from~\cite{BodlaenderCKN12} gives a significant speedup of the
dynamic programming algorithm, and it is preferable to use the reduction step only
for large tables.

Our software is publicly available, can be used under a GNU Lesser General Public Licence,
and can be downloaded at:
\begin{verse}
http://www.staff.science.uu.nl/$\sim$bodla101/java/steiner.zip
\end{verse}

This paper is organized as follows.
Some preliminary definitions are given in Section~\ref{prelim}. 
In Section~\ref{theory}, we describe both the classic dynamic
programming algorithm for Steiner
Tree on nice tree decompositions, as well as the improvement with the
 the rank-based approach as presented in \cite{BodlaenderCKN12}. 
In Section~\ref{section:experiments}, we describe the setup of our experiments,
and in Section~\ref{section:results}, we discuss the results of the experiments.
Some final conclusions are made in Section~\ref{section:conclusions}.

\section{Preliminaries}\label{prelim}
We use standard graph theory notation and quite some additional notation from~\cite{BodlaenderCKN12}. 
For a subset of edges $X \subseteq E$ of an undirected graph $G = (V,E)$,
we let $G[X]$ denote the subgraph induced by edges and endpoints of $X$, i.e. $G[X]=(V(X),X)$. We let \texttt{cc}($G$) denote the number of connected components in a graph $G$.
For a function $s$ we the function $s\setminus\{(v,s(v))\}\cup \{v,\alpha\}$ as $s[v \rightarrow \alpha]$. We use $s_{|X}$ to denote the function obtained by restricting the domain to $X$.

Given a base set $U$, we use $\mathrm{\Pi}(U)$ for the set of all partitions of $U$. It is known that, together with the coarsening relation $\sqsubseteq$, $\mathrm{\Pi}(U)$ gives the partition lattice, with the minimum element being $\{U\}$ and the maximum element being the partition into singletons. We denote $\sqcap$ for the meet operation and $\sqcup$ for the join operation in this lattice; these operators are associative and commutative. Given a partition $p \in \mathrm{\Pi}(U)$ we let \texttt{\#blocks}($p$) denote the number of blocks of $p$. If $X \subseteq U$ we let $p_{\downarrow X}\in\mathrm{\Pi}(X)$ be the partition obtained by removing all elements not in $X$ from it, and analogously we let for $U\subseteq X$ denote $p_{\uparrow X}\in \mathrm{\Pi}(X)$ for the partition obtained by adding singletons for every element in $X \setminus U$ to $p$. Also, for $X\subseteq U$, we let $U[X]$ be the partition of $U$ where one block is $\{X\}$ and all other blocks are singletons. If $a, b \in U$ we shorthand $U[ab] = U[\{a,b\}]$. The empty set, vector and partition are all denoted by $\emptyset$. 

\begin{definition}[Tree decomposition, \cite{RobertsonS2}]
A \emph{tree decomposition}  of a graph $G$ is a tree $\mathbb{T}$ in which each node $x$ has an assigned set of vertices $B_{x} \subseteq V$ (called a \emph{bag}) such that $\bigcup_{x\in\mathbb{T}}B_{x} = V$ with the following properties:
\begin{itemize}
\item for any $e = (u,v) \in E$, there exists an $x\in\mathbb{T}$ such that $u,v \in B_{x}$.
\item if $v \in B_{x}$ and $v \in B_{y}$, then $v \in B_{z}$ for all $z$ on the (unique) path from $x$ to $y$ in $\mathbb{T}$.
\end{itemize}
\end{definition}

The treewidth $tw(\mathbb{T})$ of a tree decomposition $\mathbb{T}$ is the size of the largest bag of $\mathbb{T}$ minus one, and the treewidth of a graph $G$ is the minimum treewidth over all possible tree decompositions of $G$.
\begin{definition}[Nice tree decomposition]\label{ntd} A \emph{nice tree decomposition}  is a tree decomposition with one special bag $z$ called the \emph{root} and in which each bag is one of the following types:
\begin{itemize}
\item \emph{leaf bag}: a leaf $x$ of $\mathbb{T}$ with $B_{x} = \emptyset$.
\item \emph{introduce vertex bag}: an internal vertex $x$ of $\mathbb{T}$ with one child vertex $y$ for which $B_{x} = B_{y}\cup \{v\}$ for some $v\notin B_{y}$. This bag is said to \emph{introduce} $v$.
\item \emph{introduce edge bag}: an internal vertex $x$ of $\mathbb{T}$ labelled with an edge $e = (u, v) \in E$ with one child bag $y$ for which $u, v \in B_{x} = B_{y}$. This bag is said to \emph{introduce} $e$.
\item \emph{forget vertex bag}: an internal vertex $x$ of $\mathbb{T}$ with one child bag $y$ for which $B_{x} = B_{y} \setminus\{v\}$ for some $v\in B_{y}$. This bag is said to \emph{forget} $v$.
\item \emph{join bag}: an internal vertex $x$ with two child vertices $y$ and $z$ with $B_{x} = B_{y} = B_{z}$.
\end{itemize}
We additionally require that every edge in $E$ is introduced exactly once.
\end{definition}

Nice tree decompositions were introduced in the 1990s by Kloks~\cite{Kloks93}. We use here a more
recent version that distinguishes {\em introduce edge} and {\em introduce vertex} bags \cite{CyganNPPRW11}.

Finding a tree decomposition of minimum treewidth is $\mathcal{NP}$-hard \cite{ArnborgCP87},
but for each fixed integer $k$ there is a linear time algorithm for finding a tree decomposition of width at most $k$ (if it exists) \cite{Bodlaender96}.
There are also many heuristics for finding a tree decomposition of small width; see \cite{BodlaenderK10} for
a recent overview.
Given a tree decomposition $\mathbb{T}$ of $G$, a nice tree decomposition rooted at a forget bag can be 
computed in $n \cdot \texttt{tw}^{\mathcal{O}(1)}$ time by following the arguments given in \cite{Kloks93}, with the
following modification: between a forget bag $X_i$ where we 'forget vertex $v$'  and its child bag $X_j$, we add a series of introduce edge bags for each edge $e=\{v,w\}\in E$
and $w\in X_j$.
In the remainder of the paper, we assume that a nice tree decomposition of the input graph with the appropriate width is given.
(In our experiments, we find a tree decomposition with the greedy degree heuristic, and then transform it, as discussed above, to a nice
tree decomposition of the same width.)

\section{Dynamic Programming Algorithms for Steiner Tree Parameterized by Treewidth}\label{theory}
In this section we describe both the classic dynamic programming algorithm on (nice) tree
decompositions for (the edge weighted version of)
{\sc Steiner Tree} and its variant with the rank-based approach. We first give the formal definition of the {\sc Steiner Tree} problem.

\smallskip

\noindent
\fbox{
\parbox{\textwidth}{
	\textsc{Steiner Tree}\\
	\textbf{Input:} A graph $G = (V,E)$, weight function $\omega: E \rightarrow \mathbb{N} \setminus \{0\}$, a terminal set $K \subseteq V$ and a nice tree decomposition $\mathbb{T}$ of $G$ of width \texttt{tw}.\\
	\textbf{Question:} The minimum of $\omega (X)$ over all subsets $X \subseteq E$ of $G$ such that $G[X]$ is connected and $K \subseteq V(G[X])$.
}
}

\subsection{Classic Dynamic Programming}\label{fldp}
We follow the description from \cite{BodlaenderCKN12}. In order to facilitate the correctness proof and the description of the algorithms, a collection of operators was introduced in \cite{BodlaenderCKN12}.
We will also use these operators here, and thus obtain compact descriptions of the recurrences (for the different types of nodes in the
nice tree decomposition) that shape the dynamic programming algorithm.

For each node in the nice tree decomposition, we compute a table. Each table entry maps a partition of a subset of the bag to an integer value. 
We now will introduce the notation, and give the corresponding recurrences (with just brief sketches for their correctness).

We will denote the weighted partition tables as $A_{x}(\cdot)$ and sets of partial solutions as $\mathcal{E}_{x}(\cdot)$ where $x$ denotes the current bag. 
The classic dynamic programming algorithm computes for each bag $x$ the function $A_{x}$. This function is stored in a table, with only trivial
entries (e.g., partitions mapping to infinity, as there are no forests corresponding to the partition) not stored.

We use $s: B_{x}\rightarrow \{0,1\}$ to describe which vertices belong to the tree, where $s(v) = 1$ denotes $v$ is in the Steiner tree.
For a bag $x$ and $\mathrm{s} \in \{0,1\}^{B_{x}}$ define:
\[
A_{x}(\mathrm{s})= \left\{ \left( p, \min_{X\in \mathcal{E}_{x}(p, \mathrm{s})} \omega(X) \right) \middle| p \in \mathrm{\Pi}(s^{-1}(1)) \wedge \mathcal{E}_{x}(p, \mathrm{s}) \neq \emptyset  \right\}
\]
\[
\begin{split}
\mathcal{E}_{x}(p, \mathrm{s} )  &= \Big\{ X \subseteq E_{x} \Big| \mbox{ }\forall v \in B_{x} : v \in V(G[X]) \vee v \in K \rightarrow s(v) = 1   \\
&\wedge \forall v_{1},v_{2} \in (s^{-1}(1)) : \\
& \quad v_{1},v_{2} \mbox{ are in same block in } p \leftrightarrow v_{1},v_{2} \mbox{ connected in } G[X] \\
&\wedge \mbox{\texttt{\#blocks}}(p) = \mbox{\texttt{cc}}(G[X])\Big\}
\end{split}
\]

In the definition of $\mathcal{E}_{x}(p,\mathrm{s})$,
partial solutions have (a subset of) $s^{-1}(1)$ as incident vertices in $B_{x}$ connected according to the partition $p$. The blocks in $p$ represent connected components in the partial solution. When two vertices are in the same block they belong to
the same connected component. Naturally,
any terminal $v \in K$ has to be used in a partial solution and we are allowed to use other vertices to connect these terminals. Connected components are formed by using subsets $X \subseteq E_{x}$ of edges and multiple different subsets are possible to form the same partition in a partial solution. In the partial solution tables $A_{x}(\mathrm{s})$ we only consider minimum weight partial solutions and discard any other partial solutions that are dominated. If we have a tree decomposition $\mathbb{T}$ such that its root is a forget vertex bag for $v \in K$ as input for \textsc{Steiner Tree}, then $x$ has one child $y$ with one entry in $A_{y}(\mathrm{s})$ where $s(v) = 1$. There are no other vertices in this bag since the root bag is empty.
Therefore $A_{y}(\mathrm{s})$ only contains the partition $p = \{ \{ v \} \}$ in which the single block represents a single connected component containing all terminals with minimum weight over all possible subsets of edges, thus yielding the solution.

We proceed with the recurrence for $A_{x}(\mathrm{s})$ which is used by the classic dynamic programming algorithm. In order to simplify the notation, let $v$ denote the vertex introduced and contained in an introduce bag, and let $y, z$ denote the left and right children of $x$ in $\mathbb{T}$, if present. We let $U$ respectively $U'$ denote the base set of vertices present in $y$ and $z$. We distinguish on the type of bag in $\mathbb{T}$. For a leaf bag $x$ let:
\[
A_{x}( \emptyset ) = \big\{(\emptyset, 0) \big\}
\]

This is the trivial case, where $\mathcal{E}_{x}(p, \mathrm{s})$ only contains the empty set, which doesn't contain or connect any vertices and has weight 0. 

For an introduce vertex $v$ bag $x$ with child $y$ let:
\[
A_{x}(\mathrm{s}) = \left\{ \begin{array}{l l} 
\big\{(p_{\uparrow U \cup \{v\}}, w) \big| (p,w) \in A_{y}(\mathrm{s}_{|B_{y}}) \big\}, & \quad \mbox{if } s(v)=1 \\
A_{y}(\mathrm{s}_{|B_{y}}), & \quad \mbox{if } s(v) = 0 \wedge v \notin K \\
\emptyset, & \quad \mbox{if } s(v) = 0 \wedge v \in K \\
\end{array}\right.
\]

For each partial solution in $A_{y}(\mathrm{s})$ we consider whether or not to use $v$ and add both cases (when feasible) to $A_{x}$ to fill our table for introduce vertex bag $x$. Using $v$ corresponds to $s(v)=1$, and because $v$ was just introduced and thus is currently an isolated vertex,
we insert it as a singleton into each partition. 
If we do not use $v$, i.e, $s(v)=0$ then we don't insert $v$ and preserve the same partial solution as in the child bag. If $v$ is a terminal, then not inserting $v$ is not feasible. 

For a forget vertex $v$ bag $x$ with child $y$ let:
\begin{align*}
A_{x}(\mathrm{s}) & = \texttt{rmc}(A_{y}(\mathrm{s}[v \rightarrow 0]) 
\cup \Big\{ (p_{\downarrow \overline{v}},w) \Big| (p, w) \in A_{y}(\mathrm{s}[v \rightarrow 1]) \\
&\wedge \mbox{ }\exists v' \in \overline{v}: p \sqsubseteq U[vv'] \Big\}) \text{, where}\\
\texttt{rmc}(A) &= \big\{ (p,w) \in A \big| \nexists (p, w') \in A \wedge w' < w \big\}
\end{align*}

We assume that $x$ is not the root.
The procedure basically does two steps: if $v$ is forgotten, then any partition in which $v$ is used and is a singleton gives more than one connected component. (Recall here
that the root bag forgets a terminal, and here $v$ cannot be connected to that terminal vertex.) All such entries are deleted. All other entries are `projected', i.e.,
$v$ is removed from the partitions. Possibly, multiples entries have the same projection; then we keep the one with the smallest value. The function  \texttt{rmc}($A$) expresses this.


 For an introduce edge $e = (u,v)$ bag $x$ with child $y$ let:
\[
A_{x}(\mathrm{s})= \left\{ \begin{array}{l l} 
A_{y}(\mathrm{s}), & \quad \mbox{if } s(u) = 0 \vee s(v) = 0 \\
\begin{split}
\texttt{rmc}(A_{y}(\mathrm{s}) \cup &\big\{ (\hat{U}[uv] \sqcap p_{\uparrow \hat{U}},w+\omega(u,v)) \\
& \big| (p, w) \in A_{y}(\mathrm{s}) \\
& \wedge \hat{U} = U\cup{u,v} \big\} ), 
\end{split}
& \quad \mbox{otherwise.} \\
\end{array} \right.
\]

For each partial solution in $A_{y}(\mathrm{s})$ we consider whether or not to include $e$ and add both cases (when feasible) to $A_{x}$ to fill our table for introduce edge bag $x$. If we include an edge in a partial solution then we must ensure that $u$ and $v$ are used in the partition i.e. $s(u) = s(v) = 1$. Including the edge increases the weight of the partial solution by $\omega(u,v)$ and connects the connected components containing $v$ respectively $u$,
and thus, we combine their blocks in the new partial solution. Again, if we do not include $e$, the partial solution remains the same. Because $v$ and $u$ may already have been part of the same connected component we must eliminate dominated partial solutions.

For join bag $x$ with children $y$ and $z$ let:
\[
\begin{split}
A_{x}(\mathrm{s}) = \texttt{rmc}(\Big\{ (p_{\uparrow\hat{U}}\sqcap q_{\uparrow\hat{U}}, w_{1} + w_{2}) &\Big| (p, w_{1}) \in A_{y}(\mathrm{s}) \\
&\wedge (q, w_{2}) \in A_{z}(\mathrm{s}) \wedge \hat{U} = U \cup U'     \Big\})
\end{split}
\]

Here we combine choices previously made in the subtree of $y$ with choices made in the subtree of $z$, by combining pairs of partial solutions. We account for the weight by adding their respective weights. Using edges from both partial solutions may merge connected components, so we join their connectivity. This may again result in multiple partitions of different weight, of which
we keep the minimum weight. This concludes the formulation of the recurrence for the classic dynamic programming algorithm.

The algorithm now can be expressed as follows: in bottom-up order for each bag $x$ we compute $A_x$,
and finally computes the minimum weight of a Steiner Tree by inspection the information for the root bag, as discussed above.

\subsection{Rank-Based Table Reductions}
In this section, we describe the rank-based approach from \cite{BodlaenderCKN12}. The main idea is that after we have computed a table for a bag in the nice tree
decomposition, we can carry out a reduction step and possibly remove a number of entries from the table without affecting optimality. 
A table is transformed thus to a (possibly smaller) table whose weighted partitions are {\em representative} for the collection of weighted partitions in the earlier table.
If a set of partitions extends to an optimal solution then we should also be able to extend to an optimal solution from the representative set. Representation is formally defined as:
\begin{definition}[Representation]
Given a set of weighted partitions $\mathcal{A} \subseteq \mathrm{\Pi} \times \mathbb{N}$ and a partition $q \in \mathrm{\Pi}(U)$, define:
\[
\texttt{opt}(q,\mathcal{A}) = \min \{ w | (p,w) \in \mathcal{A} \wedge p \sqcap q = \{ U \} \}
\]
For another set of weighted partitions $\mathcal{A}' \subseteq \mathrm{\Pi}(U) \times \mathbb{N}$, we say that $\mathcal{A}'$ \emph{represents} $\mathcal{A}$ if for all $q\in \mathrm{\Pi}(U)$ it holds that \texttt{opt}$(q, \mathcal{A}') =$ \texttt{opt} $(q, \mathcal{A})$.
\end{definition}

Although this definition is symmetric, we will only be interested in finding $\mathcal{A}'$ where $\mathcal{A'}\subseteq\mathcal{A}$ and where we have a size guarantee such that $\mathcal{A}'$ is small.
Omitting the formal proof (see \cite{BodlaenderCKN12}), we now state that the functions describing the formulation of the recurrence in Section \ref{fldp} preserve representation:
\begin{definition}[Preserving representation]
A function $f:2^{\mathrm{\Pi}(U)\times\mathbb{N}} \times Z \rightarrow 2^{\mathrm{\Pi}(U')\times\mathbb{N}}$ is said to \emph{preserve representation} if for every $\mathcal{A},\mathcal{A}'\subseteq\mathrm{\Pi}(U)\times\mathbb{N}$ and $z\in Z$ it holds that if $\mathcal{A}'$ represents $\mathcal{A}$ then $f(\mathcal{A}',z)$ represents $f(\mathcal{A},z)$, where $Z$ stands for any combination of further inputs.
\end{definition}

At the core of the rank-based approach, the key to obtaining a small representative set is to find for partial solutions $q \in \mathcal{A}$ the minimum weight of partial solutions $(p, w)$ such that $p \sqcap q = \{U\}$. So if we can find a set cover of partitions $p$ with minimum weight for every $q$ with this property, then we have a representative set, since when they can all extend to the unit partition, then one must also extend to the optimal solution. We can achieve this by finding a basis of minimum weight in the matrix $\mathcal{M} \in \mathbb{Z}_{2}^{ \mathrm{\Pi }(U)\times\mathrm{\Pi}(U)}$ where $\mathcal{M}[p,q] = 1$ if $p \sqcap q = \{U\}$ and $\mathcal{M}[p,q] = 0$ otherwise. In arithmetic modulo two we can rewrite this matrix as a product of two cut-matrices $\mathcal{C}$ defined as:
\begin{definition}
Define \texttt{cuts}$(t) := \{ (V_{1},V_{2}) | V_{1} \cup V_{2} = U \wedge 1 \in V_{1} \}$, where $1$ stands for an arbitrary but fixed element of $U$. Define $\mathcal{C}\in\mathbb{Z}_{2}^{\mathrm{\Pi}(U)\times\texttt{cuts}(t)}$ by $\mathcal{C} [p, (V_{1}, V_{2})] = [(V_{1},V_{2}) \sqsubseteq p]$.
\end{definition}

We now can see that $\mathcal{M} \equiv \mathcal{CC}^{T}$ and because of linear dependencies we are allowed to use the lightest (i.e., with minimal weights)
basis of the cut-matrix $\mathcal{C}$ as the representative subset $\mathcal{A'} \subseteq \mathcal{A}$ where $\mathcal{A'} \leq 2^{|U|}$. We can find this basis via straightforward Gaussian elimination in $\mathcal{C}$ after we order its rows by weight.

This yields the improved algorithm for solving \textsc{Steiner Tree}: for each node in the tree of the nice tree decomposition, in bottom-up
order, we compute a table and then reduce the size of this intermediate table by the reduce algorithm\footnote{See the proof of Theorem 3.7 in the arXiv report of \cite{BodlaenderCKN12}.}. 
The computation of the table
uses the same recurrences as for $A_x$, but as inputs we use the reduced tables for the children,
i.e., we restrict the domains --- in this way, we obtain for
each node a table whose entries are `representative' for $A_x$.

We have two variants: we can choose to 
always apply the reductions, or to apply them only in some cases. Correctness follows from the analysis in \cite{BodlaenderCKN12}.
In our experiments, we consider both the case where we always apply the reduction step, and the case where we only apply it 
when $A \geq 2^{|U|}$. Both cases give the same guarantees on the size of tables and worst case upper bound on the running time, but the
actual running times in experiments differ, as we discuss in later sections.

\section{Implementation}
\label{section:experiments}
In this section, we give some details on our implementation of the algorithms described in the previous section.

We have implemented the algorithms in Java. For each of the test graphs, we used the well known (and quite simple and effective, see e.g., \cite{BodlaenderK10}) {\em Greedy Degree} 
heuristic to find a tree decomposition. These tree decompositions were subsequently transformed into nice tree decompositions,
using the procedure which was previously described in Section~\ref{prelim}. The algorithms were executed on the thus obtained nice tree decompositions.

The recursions for the different types of nodes were implemented such that we spend linear time per generated entry (before removing double entries, and before
the reduction step). For most types, this is trivial. The computation for join bags contains a step, where we are given two partitions, and must compute
the partition that is the closure of the combination of the two (i.e., the finest partition that is a coarsening of both). We implemented this step with a
breadth first search on the vertices in the bag, with the children of a vertex $v$ all not yet discovered vertices that are in the same block as $v$ in either of the
partitions.

In order to find duplicate partial solutions we have represented the partial solution tables in a nested hash-map structure.
First we use sets of vertices that where not used in a partial solution as keys, pointing to tables of weighted partitions, effectively grouping partitions consisting of the same base set of vertices together.
These weighted partition tables are then represented by another hash-map where the partitions, which are represented as nested sets, are used as keys, pointing to the minimum weight corresponding to the partial solution. This allows us to find and replace any duplicate partial solution in amortized constant time. Java provides hash-codes for sets by adding the hash-codes for all objects contained within a set, which works well enough for the outer hash-table used in our structure. This standard approach breaks down when we use it to calculate hash-codes for partitions however, at it effectively adds all hash-codes of vertices used in the partition together. This results in the same hash-code for all partitions used in the same inner hash-map. 
To resolve this problem we disrupt this commutative effect by multiplying indexes of vertices contained in each block, and then taking the sum of these values of
blocks in order to calculate hash-codes for partitions. We apply the multiplications modulo a prime number to avoid integer overflows.
In our experiments, we observed that this approach results in approximately 3\% collisions for large tables.

In the implementation of the rank-based approach we receive partial solution tables from the classic algorithm.
For every computed partial solution table we enumerate the cuts and fill the cut-matrix ordering its rows by weight. For every partial solution represented in the matrix we then find the leading 1 in its row, after which we add the values in its row to the row of every other partial solution of higher weight containing a 1 in the same column, modulo 2. We then include the partial solution in $\mathcal{A'}$. Any time we find a solution with a row consisting of all 0's we can eliminate it, as it is linearly dependent on previously processed partial solutions. We can stop when all partial solutions have been processed, or when we have processed $2^{|U|}$ rows, since all remaining partial solutions are linearly dependent on solutions in $\mathcal{A}$. Any time a partial solution is processed we can eliminate the column containing its leading 1, since all elements in this column are 0.

Chimani et al.~\cite{ChimaniMZ12} give an efficient algorithm for Steiner tree for graphs given with a tree decomposition,
that runs in time $O(B^2_{k+2} k n)$ time, with $k$ the width of the tree decomposition. We have chosen not to use the
coloring scheme from Chimani et al.~\cite{ChimaniMZ12}, but instead use hash tables (as discussed above) to store the 
tables. Of course, our choice has the disadvantage that we lose a guarantee on the worst case running time (as we cannot
rule out scenarios where many elements are hashed to the same position in the hash table), but gives a simple mechanism
which works in practice very well. In fact, if we assume that the expected number of collisions of an element in the hash
table is bounded by a constant (which can be observed in practice), then the expected running time of our implementation
matches asymptotically  the worst case running time of Chimani et al.

\section{Experimental Results}
\label{section:results}
In this section, we will report the results for experiments with the algorithms discussed in Section~\ref{theory}. We will denote the classic dynamic programming algorithm as CDP. 
With RBA, we denote the algorithm where we always apply the reduction step,
whereas RBC denotes the algorithm which only applies the reduction step when we have a table whose size is larger than the bound guaranteed by reduction.
We will compare the runtime of these three algorithms. 
Furthermore we will compare the number of partial solutions generated during the execution of these algorithms to illustrate how much work is being saved by reducing the tables.

Each of the three algorithms receives as input the same nice tree decomposition of the input graph; this nice tree decomposition is rooted at a forget bag of a terminal vertex.
The experiments where performed on sets of graphs of different origin, spanning a range of treewidth sizes of their tree decompositions, and where possible diversified on the number of vertices, edges and terminals. 
Our graphs come from benchmarks for algorithms for the {\sc Steiner Tree} problem and for Treewidth. The graphs from Steiner tree benchmarks can be
found in Steinlib \cite{KochMV}, a repository for Steiner Tree problems. These are prefixed by \emph{b}, \emph{i080} or \emph{es}.
Graph instances prefixed by \emph{b} are randomly generated sparse graphs with edge weights between 1 and 10; these were introduced in \cite{Beasley84} and were generated following a scheme outlined in \cite{Aneja80}.
The \emph{i080} graph instances are randomly generated sparse graphs with incidence edge weights, introduced in \cite{Duin93}. We have grouped these sparse graphs together in the results. The next set of instances, prefixed by \emph{es}, were generated by placing random points on a two-dimensional grid, which serve as terminals. By building the grid outlined in \cite{Hanan66} they where converted to rectilinear graphs with L1 edge weights and preprocessed with GeoSteiner \cite{WarmeWZ}.
The last collection of graphs come are often used as benchmarks for algorithms for {\sc Treewidth}. These
come from 
from Bayesian network and graph colouring applications. 
We transformed these to {\sc Steiner Tree} instances by adding random edge weights between 1 and 1000, and by selecting randomly a
subset of the vertices as terminals (about 20\% of the original vertices). These graphs can be found in \cite{treewidthlib}.

All algorithms have been implemented in Java and the computations have been carried out on a Windows-7 operated PC with an Intel Core i5-3550 processor and 16.0 GB of available main memory. 
We have given each of the algorithms a maximum time of one hour to find a solution for a given instance; in the tables,
we marked instances halted due to the use of the maximum time by a *.

\begin{table}[!h]
\begin{center}
\begin{tabular}{l l l l l | l l l }
\hline
instance & \texttt{tw}($\mathbb{T}$) & $|V|$ & $|E|$ & $|K|$ & CDP & RBA & RBC\\
\hline
\hline
b01.stp	&	4	&	50	&	63	&	9	&	55	&	53	&	17	\\
b02.stp	&	4	&	50	&	63	&	13	&	12	&	30	&	12	\\
b08.stp	&	6	&	75	&	94	&	19	&	171	&	92	&	48	\\
b09.stp	&	6	&	75	&	94	&	38	&	78	&	46	&	31	\\
b13.stp	&	7	&	100	&	125	&	17	&	1328	&	618	&	408	\\
b14.stp	&	7	&	100	&	125	&	25	&	2190	&	385	&	275	\\
b15.stp	&	8	&	100	&	125	&	50	&	14421	&	1542	&	1281	\\
i080-001.stp	&	9	&	80	&	120	&	6	&	98617	&	11270	&	7953	\\
i080-003.stp	&	9	&	80	&	120	&	6	&	144796	&	12689	&	10211	\\
i080-004.stp	&	10	&	80	&	120	&	6	&	1618531	&	70192	&	68930	\\
b06.stp	&	10	&	50	&	100	&	25	&	1325669	&	36986	&	29082	\\
b05.stp	&	11	&	50	&	100	&	13	&	*	&	270376	&	207516\\
i080-005.stp &	11	&	80	&	120	&	6	&	*	&	936074	&	840466\\
\hline
\end{tabular}
\end{center}
\caption{Runtime in milliseconds for instances from Steinlib (1)}
\label{t1:a}
\end{table}

\begin{table}[!h]
\begin{center}
\begin{tabular}{l l l l l | l l l }
\hline
instance & \texttt{tw}($\mathbb{T}$) & $|V|$ & $|E|$ & $|T|$ & CDP & RBA & RBC\\
\hline
\hline
es90fst12.stp	&	5	&	207	&	284	&	90	&	71	&	120	&	60	\\
es100fst10.stp	&	5	&	229	&	312	&	100	&	105	&	166	&	86	\\
es80fst06.stp	&	6	&	172	&	224	&	80	&	272	&	276	&	151	\\
es100fst14.stp	&	6	&	198	&	253	&	100	&	109	&	160	&	78	\\
es90fst01.stp	&	7	&	181	&	231	&	90	&	250	&	270	&	148	\\
es100fst13.stp	&	7	&	254	&	361	&	100	&	1223	&	1200	&	679	\\
es100fst15.stp	&	8	&	231	&	319	&	100	&	2600	&	1688	&	1033	\\
es250fst03.stp	&	8	&	543	&	727	&	250	&	2904	&	2010	&	1251	\\
es100fst08.stp	&	9	&	210	&	276	&	100	&	4670	&	2302	&	1942	\\
es250fst05.stp	&	9	&	596	&	832	&	250	&	24460	&	15550	&	9742	\\
es250fst07.stp	&	10	&	585	&	799	&	250	&	107150	&	54605	&	31729	\\
es500fst05.stp	&	10	&	1172	&	1627	&	500	&	124664	&	47336	&	31102	\\
es250fst12.stp	&	11	&	619 	&	872	&	250	&	*	&	144932 & 	95855\\
es100fst02.stp	&	12	&	339	&	522	&	100	&	*	&	426078 &	334785\\
es250fst01.stp	&	12	&	623	&	876	&	250	&	*	&	332389 & 	246704\\
es250fst08.stp	&	13	&	657	&	947	&	250	&	*	&	2670464 &	2251728\\
es250fst15.stp	&	13	&	713	&	1053	&	250	&	*	&	2120913 &	1671672\\
\hline
\end{tabular}
\end{center}
\caption{Runtime in milliseconds for instances from Steinlib (2)}
\label{t1:b}
\end{table}

\begin{table}[!h]
\begin{center}
\begin{tabular}{l l l l l | l l l }
\hline
instance & \texttt{tw}($\mathbb{T}$) & $|V|$ & $|E|$ & $|T|$ & CDP & RBA & RBC\\
\hline
\hline
myciel3.stp	&	5	&	11	&	20	&	2	&	5	&	8	&	4	\\
BN\_28.stp	&	5	&	24	&	49	&	4	&	8	&	15	&	7	\\
pathfinder.stp	&	6	&	109	&	211	&	21	&	422	&	254	&	155	\\
csf.stp	&	6	&	32	&	94	&	6	&	335	&	198	&	116	\\
oow-trad.stp	&	7	&	33	&	72	&	6	&	766	&	594	&	364	\\
mainuk.stp	&	7	&	48	&	198	&	9	&	8842	&	3495	&	2025	\\
ship-ship.stp	& 	8	&	50	&	114	&	10	&	10579	&	4511	&	2841	\\
barley.stp	&	8	&	48	&	126	&	9	&	9281	&	2410	&	1473	\\
miles250.stp	&	9	&	128	&	387	&	25	&	35369	&	14423	&	9382	\\
jean.stp	&	9	&	80	&	254	&	16	&	39192	&	18237	&	10862	\\
huck.stp	&	10	&	74	&	301	&	14	&	\textit{17030}	&	38486	&	21050	\\
myciel4.stp	&	11	&	23	&	71	&	4	&	1510595	&	98720		&	93107 \\
munin1.stp	&	11	&	189	&	366	&	37	&	*	&	460051	&	372718 \\
pigs.stp	&	12	&	441	&	806	&	88	&	*	&	1431083	&	1280194 \\
anna.stp	&	12	&	138	&	493	&	27	&	*	&	*	&	3291591 \\
\hline
\end{tabular}
\end{center}
\caption{Runtime in milliseconds for instances on graphs from TreewidthLib}
\label{t1:c}
\end{table}

In Tables \ref{t1:a} -- \ref{t1:c}, we have gathered the results for the runtimes of the three algorithms for the aforementioned graph instances. 
We immediately notice that RBC outperforms RBA in all cases.
If we investigate Tables \ref{t2:a} -- \ref{t2:c} we notice
that the number of partial solutions computed during RBA is not significantly smaller compared to the number computed during RBC. From these results we can conclude that it is preferable to use the reductions more sparingly in order to decrease runtime, since applying the reductions when the tables are already smaller than their size guarantee does not seem to have a noteworthy effect.

We also notice that, while RBA outperforms CDP in numerous cases, RBC outperforms CDP in all but one (discussed below). 
For example, in the case of \emph{i080-004} we see a significant speed-up: the classic DP uses 26 minutes to find the optimal solution, but 
RBC uses just 69 seconds.
Furthermore we see a strong increase in the runtime difference when the width of the tree decompositions increases. 
This is further reflected in Table \ref{t2:a} -- \ref{t2:c} where we see that when the width of the tree 
decompositions increases, the difference in the number of of generated partial solutions grows significantly.

\begin{table}[!h]

\centering{
\begin{tabular}{l l l l l | l l l }
\hline
instance & \texttt{tw}($\mathbb{T}$) & $|V|$ & $|E|$ & $|T|$ & CDP & RBA & RBC\\
\hline
\hline
b01.stp	&	4	&	50	&	63	&	9	&	3141	&	2854	&	2854	\\
b02.stp	&	4	&	50	&	63	&	13	&	3263	&	2763	&	2769	\\
b08.stp	&	6	&	75	&	94	&	19	&	39178	&	11278	&	11345	\\
b09.stp	&	6	&	75	&	94	&	38	&	18970	&	5177	&	5449	\\
b13.stp	&	7	&	100	&	125	&	17	&	328366	&	68533	&	70693	\\
b14.stp	&	7	&	100	&	125	&	25	&	400940	&	35554	&	40012	\\
b15.stp	&	8	&	100	&	125	&	50	&	2294557	&	84567	&	94951	\\
i080-001.stp	&	9	&	80	&	120	&	6	&	15757284	&	529805	&	565777	\\
i080-003.stp	&	9	&	80	&	120	&	6	&	18841974	&	589313	&	589773	\\
i080-004.stp	&	10	&	80	&	120	&	6	&	196513167	&	2611426	&	3270334	\\
b06.stp	&	10	&	50	&	100	&	25	&	156669926	&	903700	&	938800	\\
b05.stp	&	11	&	50	&	100	&	13	&	*	&	6320072	&	6320264	\\
i080-005.stp &	11	&	80	&	120	&	6	&	*	&	26653282	&	31275766	\\
\hline
\end{tabular}
}
\caption{Number of generated partial solutions for instances of Steinlib (1)}
\label{t2:a}
\end{table}

\begin{table}[!h]
\centering{
\begin{tabular}{l l l l l | l l l }
\hline
instance & \texttt{tw}($\mathbb{T}$) & $|V|$ & $|E|$ & $|T|$ & CDP & RBA & RBC\\
\hline
\hline
es90fst12.stp	&	5	&	207	&	284	&	90	&	39324	&	28739	&	28761	\\
es100fst10.stp	&	5	&	229	&	312	&	100	&	53477	&	36548	&	36578	\\
es80fst06.stp	&	6	&	172	&	224	&	80	&	97373	&	47524	&	48124	\\
es100fst14.stp	&	6	&	198	&	253	&	100	&	51786	&	34747	&	34792	\\
es90fst01.stp	&	7	&	181	&	231	&	90	&	83763	&	36783	&	36825	\\
es100fst13.stp	&	7	&	254	&	361	&	100	&	364446	&	138376	&	138635	\\
es100fst15.stp	&	8	&	231	&	319	&	100	&	596847	&	163386	&	163410	\\
es250fst03.stp	&	8	&	543	&	727	&	250	&	700715	&	210195	&	210319	\\
es100fst08.stp	&	9	&	210	&	276	&	100	&	825150	&	108249	&	113399	\\
es250fst05.stp	&	9	&	596	&	832	&	250	&	4708395	&	921183	&	922000	\\
es250fst07.stp	&	10	&	585	&	799	&	250	&	17267208	&	2106090	&	2107053	\\
es500fst05.stp	&	10	&	1172	&	1627	&	500	&	19211081	&	2263151	&	2263681	\\
es250fst12.stp	&	11	&	619 	&	872	&	250	&	*	&	4641299	&	4642325 \\
es100fst02.stp	&	12	&	339	&	522	&	100	&	*	&	5531945	&	5532151 \\
es250fst01.stp	&	12	&	623	&	876	&	250	&	*	&	5079895	&	5080613 \\
es250fst08.stp	&	13	&	657	&	947	&	250	&	*	&	20876551	&	21907601 \\
es250fst15.stp	&	13	&	713	&	1053	&	250	&	*	&	17467070	&	17698575 \\
\hline
\end{tabular}
}
\caption{Number of generated partial solutions for instances of Steinlib (2)}
\label{t2:b}
\end{table}

\begin{table}[!h]
\centering{
\begin{tabular}{l l l l l | l l l }
\hline
instance & \texttt{tw}($\mathbb{T}$) & $|V|$ & $|E|$ & $|T|$ & CDP & RBA & RBC\\
\hline
\hline
myciel3.stp	&	5	&	11	&	20	&	2	&	2763	&	1773	&	1837	\\
BN\_28.stp	&	5	&	24	&	49	&	4	&	5317	&	3509	&	3529	\\
pathfinder.stp	&	6	&	109	&	211	&	21	&	130730	&	31126	&	31789	\\
csf.stp	&	6	&	32	&	94	&	6	&	104620	&	28196	&	28908	\\
oow-trad.stp	&	7	&	33	&	72	&	6	&	235555	&	66378	&	66443	\\
mainuk.stp	&	7	&	48	&	198	&	9	&	2109366	&	326069	&	330049	\\
ship-ship.stp	&	8	&	50	&	114	&	10	&	2439667	&	294814	&	295219	\\
barley.stp	&	8	&	48	&	126	&	9	&	1825048	&	168018	&	169318	\\
miles250.stp	&	9	&	128	&	387	&	25	&	8955601	&	531507	&	542261	\\
jean.stp	&	9	&	80	&	254	&	16	&	8886646	&	657659	&	690002	\\
huck.stp	&	10	&	74	&	301	&	14	&	5916126	&	1103232	&	1114182	\\
myciel4.stp	&	11	&	23	&	71	&	4	&	211076605	&	2156903	&	3798591 \\
munin1.stp	&	11	&	189	&	366	&	37	&	*	&	15486584	&	19662020 \\
pigs.stp	&	12	&	441	&	806	&	88	&	*	&	14525282	&	17868488 \\
anna.stp	&	12	&	138	&	493	&	27	&	*	&	*	&	55923467 \\
\hline
\end{tabular}
}
\caption{Number of generated partial solutions for graphs from TreewidthLib}
\label{t2:c}
\end{table}

The \emph{huck} instance is the only example where using the rank-based approach does not pay off.
Upon further inspection we found that the tree decomposition for this instance has only one bag of size 11, while most of the other bags are of size 7 and below. This is also reflected by the difference in the number
of generated partial solutions, where the improvement factor is not comparable to the other cases. Vice versa we found that the \emph{i080-004} case included 18 bags of treewidth 11 of which 6 where join bags, which explains the extreme difference. In practice, when we run dynamic programming algorithms on tree decompositions, the underlying structure of the decomposition has a big influence on the performance, which is not always properly reflected by the treewidth of a graph. In general however, the rank-based approach is more and more advantageous as the
treewidth increases, even allowing us to find solutions where CDP does not
find any within the time limit. During the execution of the experiments we have also tracked the amount of time spent on filling the cut-matrices and the time spent on performing Gaussian elimination, and found that we spent significantly more time on filling the table whereas the Gaussian elimination step only takes up a small fraction of the time spent on reducing the tables.

\begin{table}[!h]
\centering{
\begin{tabular}{l | l l l}
\hline
bag size & CDP & RBA & RBC\\
\hline
\hline
0	&	1	&	1	&	1	\\
1	&	2	&	2	&	2	\\
2	&	4	&	4	&	4	\\
3	&	15	&	15	&	15	\\
4	&	50	&	49	&	49	\\
5	&	196	&	168	&	168	\\
6	&	782	&	541	&	541	\\
7	&	3438	&	1441	&	1441	\\
8	&	15746	&	4901	&	4901	\\
9	&	69315	&	13300	&	13300	\\
10	&	252560	&	41740	&	41740	\\
11	&	860867	&	80694	&	80694	\\
\hline
\end{tabular}
}
\caption{The maximum number of partial solutions found in bags of a fixed size during the execution of the three algorithms for the \emph{es500fst05} instance}
\label{t3}
\end{table}

To further illustrate the advantages of the rank-based approach as the
treewidth increases, we have included Table \ref{t3} which shows the effect on the maximum number of partial solutions found in bags of a fixed size in the \emph{es500fst05} instance. We notice no difference between RBA and RBC which is not unexpected when considering the previous results. We see a strong reduction in the size of the intermediate tables for bigger treewidth when compared to CDP, indicating a worthy pay-off for the time spent reducing the size of partial solution tables.

\section{Discussion and Concluding Remarks}
\label{section:conclusions}
In this paper, we presented an experimental evaluation of the rank based approach by Bodlaender et al.~\cite{BodlaenderCKN12}, comparing the classic dynamic programming for 
\textsc{Steiner Tree} and the new versions based on Gaussian elimination. The results are very promising:
even for relatively small values of the width of the tree decompositions, the new approach shows
a notable speed-up in practice.
The theoretical analysis of the algorithm already predicts that the new algorithms are asymptotically faster,
but it is good to see that the improvement already is clearly visible at small size benchmark instances.

Overall, the rank based approach is an example of the general technique of representativity: a powerful but
so far underestimated paradigmatic improvement to dynamic programming. A further exploration of this
concept, both in theory (improving the asymptotic running time for problems) as in experiment and
algorithm engineering seems highly interesting. Our current paper gives a clear indication of the practical
relevance of this concept.

We end this paper with a number of specific points for further study:
\begin{itemize}
\item The rank based approach promises also faster algorithms on tree decompositions for several other 
problems. The experimental evaluation can be executed for other problems. In particular, for
{\sc Hamiltonian Circuit} and similar problems, it would be interesting to compare the
use of the basis from \cite{BodlaenderCKN12} with  the smaller basis
given by Cygan et al.~\cite{CyganKN12}.
\item How well does the \emph{Cut and Count} method perform? 
As remarked in~\cite{CyganNPPRW11}, it seems advantageous to use polynomial identity testing rather then the isolation lemma to optimize the running time.
\item Are further significant improvements on the running time possible by using different data structures
or variants of the approach, e.g., by not storing table entries as partitions of subsets by identifying them
by their row in the matrix $\cal{M}$? 
\item Are running time improvements possible by other forms of reduction of tables (without affecting 
optimality)? If we exploit the two families theorem by Lov\'{a}sz \cite{Lovasz77},
we obtain a variant of our algorithm, with a somewhat different reduce algorithm \cite{FominLS13} (see
also \cite{Marx09}); how does the running time of this version compare with the running time
of the algorithm we studied?
\item Can we use the rank based approach to obtain a faster version of
the {\em tour merging} heuristic for TSP by Cook
and Seymour \cite{CookS03}? Also, it would be interesting to try a variant of tour merging 
for other problems, e.g., `tree merging' as a heuristic for {\sc Steiner Tree}.
\item For what other problems does the rank based approach give faster algorithms in
practical settings? 
\item Are there good heuristic ways of obtaining small representative sets, even for problems where
theory tells us that representative sets are large in the worst case?
\end{itemize}


\bibliographystyle{abbrv}
\bibliography{definitions,papers}

\end{document}